# Flow of Jeffrey Fluid through Narrow Tubes

Santhosh Nallapu, G. Radhakrishnamacharya

**Abstract** — The present paper deals with a two-fluid model for the flow of Jeffrey fluid in tubes of small diameters. It is assumed that the core region consists of Jeffrey fluid and Newtonian fluid in the peripheral region. Analytical expressions for velocity, effective viscosity, core hematocrit and mean hematocrit have been derived. The effects of various parameters, namely, Jeffrey parameter ($\lambda_1$), tube hematocrit ($H_0$) and tube radius (a) on effective viscosity, core hematocrit and mean hematocrit have been studied. It is found that the effective viscosity decreases as the Jeffrey parameter increases but increases with tube hematocrit and tube radius. Further, the core hematocrit decreases with Jeffrey parameter, tube hematocrit and tube radius. It is also noticed that the flow exhibits the anomalous Fahraeus-Lindquist effect.

—————————— ◆ ——————————

## 1 INTRODUCTION

THE microcirculation is defined as blood flow in small blood vessels, namely, arterioles, capillaries and venules, whose diameters range from 20 µm (microns) to 500 µm. Its main functions are the transport of oxygen and nutrients to cells of the body, removal of waste products such as carbon dioxide and urea, circulation of molecules and cells that mediate the organism's defence and immune response, and play a fundamental role in the tissue repair process. Some anomalous effects like Fahraeus-Lindquist effect, Fahraeus effect and existence of a cell-free or cell-depleted layer near the wall, are observed in microcirculation.

The study of blood flow in small diameter tubes is very important in physiological and clinical problems and hence has attracted considerable attention of researchers. Sharan and Popel [1] considered a two-phase model to discuss the flow of blood in narrow tubes. Haynes [2] and Bugliarello and Sevilla [3] have considered a two-fluid model with both fluids as Newtonian fluids and with different viscosities.

It is realised that blood being a suspension of corpuscles, behaves like a non-Newtonian fluid at lower shear rates (Haynes and Burton [4], Hershey and Chow [5]). Hence, several non-Newtonian fluid models have been considered for blood flow in small diameter tubes. Chaturani and Upadhya [6], [7] and Shukla et al. [8], [9] considered two and three-layered blood flow models assuming blood as a polar fluid. Srivastava and Saxena [10], Haldar and Andersson [11] have presented a two-layered blood flow model in which the core region is occupied by a Casson type fluid. Aroesty and Gross [12] analysed pulsatile flow in small vessels treating blood as a Casson fluid.

A non-Newtonian fluid model that has attracted many researchers is Jeffrey fluid model, as this is used to represent a physiological fluid (Hayat et al. [13]). Jeffrey fluid model is a significant generalization of Newtonian fluid model as the later one can be deduced as a special case of the former. Several researchers have studied about Jeffrey fluid under different conditions. Hayat et al. [14] analyzed three-dimensional flow of Jeffrey fluid. Vajravelu et al. [15] investigated the influence of heat transfer on peristaltic transport of a Jeffrey fluid. Devaki et al. [16] have considered the pulsatile flow of a Jeffrey fluid in a circular tube lined internally with porous material. Akbar et al. [17] stuided the Jeffrey fluid model for blood through a tapered artery with a stenosis. However, the flow of a Jeffrey fluid in tubes of small diameter has not received any attention.

Hence, in the present paper, a two-fluid model for the flow of Jeffrey fluid in tubes of small diameters, has been investigated. It is assumed that the core region consists of Jeffrey fluid and the peripheral layer consists of Newtonian fluid. Making the assumptions as in Chaturani and Upadhya [6], the linearised equations of motion have been solved and analytical solution has been obtained. The expressions for velocity, effective viscosity, core hematocrit and mean hematocrit have been derived for cell-free layer and the effects of relevant parameters on these flow variables have been studied.

## 2 FORMULATION OF THE PROBLEM

Let us consider a laminar, steady and axisymmetric flow of an incompressible fluid through a rigid circular tube of constant radius a. It is assumed that the flow in the tube is represented by a two-fluid model consisting of a core region of radius b, occupied by Jeffrey fluid and peripheral region of thickness (a-b=ε) filled by Newtonian fluid as shown in Fig. 1. Let $\mu_p$ be the viscosity of Newtonian fluid in the peripheral region and $\mu_c$ be the viscosity of Jeffrey fluid in the core region.

- **Corresponding Author:** *G.Radhakrishnamacharya, Professor of Mathematics, National Institute of Technology Warangal, India, E-mail: grk.nitw@yahoo.com*

- *Santhosh Nallapu, Research Scholar in anonymous Institute and Department, E-mail: princenallapu@gmail.com*

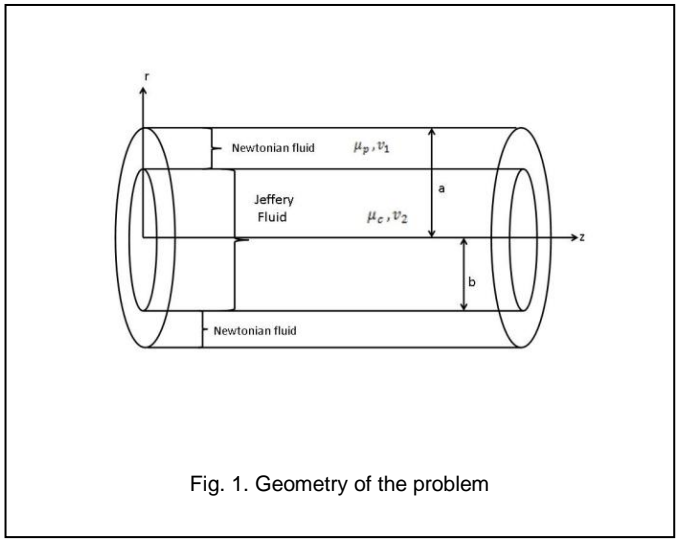

Fig. 1. Geometry of the problem

The constitutive equations for an incompressible Jeffrey fluid [13] are

$$\bar{T} = -\bar{P}\bar{I} + \bar{S}$$
$$\bar{S} = \frac{\mu}{1+\lambda_1}\left(\bar{\dot{\gamma}} + \lambda_2 \bar{\ddot{\gamma}}\right) \quad (1)$$

where $\bar{T}, \bar{S}$ are Cauchy stress tensor and extra stress tensor respectively, $P$ is the pressure, $\bar{I}$ is the identity tensor, $\lambda_1$ is the ratio of relaxation to retardation times, $\lambda_2$ is the retardation time, $\mu$ is the dynamic viscosity, $\dot{\gamma}$ is the shear rate and dots over the quantities indicate differentiation with respect to time. Cylindrical polar coordinate system (r, θ, z) is chosen where the z axis is taken along the axis of the tube.

The equations governing the steady two-dimensional flow of an incompressible Jeffrey fluid are:

Equation of continuity:

$$\frac{\partial v_r}{\partial r} + \frac{v_r}{r} + \frac{\partial v_z}{\partial z} = 0 \quad (2)$$

and

$$\rho\left[v_r\frac{\partial}{\partial r} + v_z\frac{\partial}{\partial z}\right]v_r = -\frac{\partial p}{\partial r} + \frac{1}{r}\frac{\partial}{\partial r}(r\bar{S}_{rr}) + \frac{\partial}{\partial z}(\bar{S}_{rz}) \quad (3)$$

$$\rho\left[v_r\frac{\partial}{\partial r} + v_z\frac{\partial}{\partial z}\right]v_z = -\frac{\partial p}{\partial z} + \frac{1}{r}\frac{\partial}{\partial r}(r\bar{S}_{zr}) + \frac{\partial}{\partial z}(\bar{S}_{zz}) \quad (4)$$

in which

$$\bar{S}_{rr} = \frac{2\mu_c}{1+\lambda_1}\left[1 + \lambda_2\left(v_r\frac{\partial}{\partial r} + v_z\frac{\partial}{\partial z}\right)\right]\left(\frac{\partial v_r}{\partial r}\right) \quad (5)$$

$$\bar{S}_{rz} = \bar{S}_{zr} = \frac{\mu_c}{1+\lambda_1}\left[1 + \lambda_2\left(v_r\frac{\partial}{\partial r} + v_z\frac{\partial}{\partial z}\right)\right]\left(\frac{\partial v_z}{\partial r} + \frac{\partial v_r}{\partial z}\right) \quad (6)$$

$$\bar{S}_{zz} = \frac{2\mu_c}{1+\lambda_1}\left[1 + \lambda_2\left(v_r\frac{\partial}{\partial r} + v_z\frac{\partial}{\partial z}\right)\right]\left(\frac{\partial v_z}{\partial z}\right) \quad (7)$$

where $v_r, v_z$ are the velocity components in the r and z directions respectively, p is the pressure, ρ is the density and $\bar{S}_{rr}, \bar{S}_{rz}, \bar{S}_{zr}, \bar{S}_{zz}$ are the extra stress components.

It is assumed that the flow is in z-direction only and hence the velocity component $v_r = 0$. Consequently, the equations governing the flow of fluid (Jeffrey fluid) in the core region (0≤ r ≤ b) reduce to,

$$\frac{\partial p}{\partial r} = 0 \quad (8)$$

$$\frac{\mu_c}{1+\lambda_1}\frac{1}{r}\frac{\partial}{\partial r}\left(r\frac{\partial v_z}{\partial r}\right) - \frac{\partial p}{\partial z} = 0 \quad (9)$$

Let $v_z(r) = v_1(r)$ be the velocity in the peripheral region and $v_2(r)$ in the core region. The equations governing the flow of fluid are:

Peripheral region (Newtonian fluid):

$$\mu_p \frac{1}{r}\frac{\partial}{\partial r}\left(r\frac{\partial v_1}{\partial r}\right) - \frac{\partial p}{\partial z} = 0 \quad \text{for} \ \ b \le r \le a \quad (10)$$

Core region (Jeffrey fluid):

$$\frac{\mu_c}{1+\lambda_1}\frac{1}{r}\frac{\partial}{\partial r}\left(r\frac{\partial v_2}{\partial r}\right) - \frac{\partial p}{\partial z} = 0 \quad \text{for} \ \ 0 \le r \le b \quad (11)$$

where $\frac{\partial p}{\partial z}$ is the constant pressure gradient.

The boundary conditions for the problem are given by

$$v_1 = 0 \quad \text{at} \quad r = a$$
$$v_1 = v_2, \quad \tau_1 = \tau_2 \quad \text{at} \quad r = b$$
$$v_2 \ \text{is finite at} \quad r = 0 \quad (12 \text{ a,b,c})$$

Condition (12a) is the classical no-slip boundary condition for the velocity, (12b) denotes the continuity of velocities and stresses at the interface and (12c) is the regularity condition.

Solving equations (10) and (11) under the conditions (12), we get

$$v_1(\eta) = \frac{a^2 P}{4\mu_p}(1-\eta^2) \quad \text{for,} \quad d \le \eta \le 1 \quad (13)$$

$$v_2(\eta) = \frac{a^2 P}{4\mu_p}(1-d^2 + \mu'(1+\lambda_1)(d^2-\eta^2)) \quad \text{for} \ 0 \le \eta \le d \quad (14)$$

where

$$\eta = \frac{r}{a}, \quad d = \frac{b}{a}, \quad P = -\frac{\partial p}{\partial z}, \quad \mu' = \frac{\mu_p}{\mu_c} \tag{15}$$

The flow flux in the peripheral region, denoted by $Q_P$, is defined by

$$Q_p = 2\pi a^2 \int_d^1 v_1(\eta).\eta d\eta \tag{16}$$

Substituting for $v_1(\eta)$ from (13) in (16), we get

$$Q_p = \frac{a^4 P \pi}{8\mu_p}(1 - 2d^2 + d^4) \tag{17}$$

Similarly, the flow flux in the core region is given by

$$Q_c = 2\pi a^2 \int_0^d v_2(\eta).\eta d\eta = \frac{a^4 P \pi}{8\mu_p}(2(d^2 - d^4) + \mu'(1+\lambda_1)d^4) \tag{18}$$

Thus, the flow flux through the tube is given by

$$Q = Q_p + Q_c \tag{19}$$

using (16) and (17) in (18), we get

$$Q = \frac{a^4 P \pi}{8\mu_p}(1 - d^4 + \mu'(1+\lambda_1)d^4) \tag{20}$$

Comparing (20) with flow flux for Poiseuille flow, we get the effective viscosity as

$$\mu_{eff} = \frac{\mu_p}{1 - d^4 + \mu'(1+\lambda_1)d^4} \tag{21}$$

If we take $\lambda_1 = 0$ in (21), i.e., for Newtonian fluid, it reduces to

$$\mu_{eN} = \frac{\mu_p}{1 - d^4 + \mu' d^4} \tag{22}$$

This is same as the expression obtained by Buglierello and Sevilla [3].

### 2.1 Mean Hematocrit for Cell-free Wall Layer

The percentage volume of red blood cells is called the hematocrit and is approximately 40-45% for adults.

The core hematocrit $H_c$ is related to the hematocrit $H_0$ of blood leaving or entering the tube as,

$$H_0 Q = H_c Q_c \tag{23}$$

Substituting for, $Q_c$ and $Q$ from (18) and (20) in (23), we get (after simplification),

$$\overline{H}_c = \frac{H_c}{H_0} = 1 + \frac{1 - 2d^2 + d^4}{2d^2(1-d^2) + \mu'(1+\lambda_1)d^4} \tag{24}$$

where $\overline{H}_c$ is the normalized core hematocrit.

The mean hematocrit within the tube $H_m$ is given by

$$H_m \pi a^2 = H_c \pi b^2 \tag{25}$$

Substituting for $H_c$ from equation (24), we get

$$\overline{H}_m = \frac{H_m}{H_0} = \left(1 + \frac{1 - 2d^2 + d^4}{2d^2(1-d^2) + \mu'(1+\lambda_1)d^4}\right) d^2 \tag{26}$$

where $\overline{H}_m$ is the normalized mean hematocrit.

## 3 RESULTS AND DISCUSSION

The expressions for effective viscosity $\mu_{eff}$, core hematocrit $\overline{H}_c$ and mean hematocrit $\overline{H}_m$ are given by (21), (24) and (25) respectively. The effects of various parameters, namely, Jeffrey parameter ($\lambda_1$), tube hematocrit ($H_0$) and tube radius (a) on the above characteristics have been computed using Mathematica software and the results are graphically presented in Figs. (2 - 13). The value of d (non-dimensional core radius) is calculated from the relation: d = 1 - (ε/a), in which ε denotes the peripheral layer thickness for a given hematocrit. The values of ε are 3.12 μ for 40% hematocrit, 3.60 μ for 30% and 4.67 μ for 20% ( Haynes, Table 1, [2] ).

In Figs 2 - 5, the effects of Jeffrey parameter $\lambda_1$ and tube hematocrit $H_0$ on effective viscosity $\mu_{eff}$ have been shown for several values of tube radius a. It is observed that the effective viscosity decreases as the Jeffrey parameter $\lambda_1$ increases but increases with tube hematocrit $H_0$ (Figs. 2 - 5). These results are in agreement with the results of Haynes [2], Chaturani and Upadhya [6] and Srivastava [18]. Further, for given values of Jeffrey parameter and tube hematocrit, the effective viscosity increases with tube radius a, i.e., the flow exhibits Fahraeus-Lindquist effect (i.e., apparent viscosity of blood increases with increasing tube diameter).

The effects of tube radius (a), Jeffrey parameter $\lambda_1$ and tube hematocrit $H_0$ on core hematrocit $\overline{H}_c$ and mean hematocrit $\overline{H}_m$ are depicted in Figs. 6-13. It can be seen that the core hematocrit $\overline{H}_c$ decreases with Jeffrey parameter $\lambda_1$, tube hematocrit $H_0$ and tube radius a (Figs. 6-9). Figs. 10-13 show that the mean hematocrit $\overline{H}_m$ decreases as Jeffrey parameter $\lambda_1$ increases but increases with tube hematocrit $H_0$ and tube radius a.

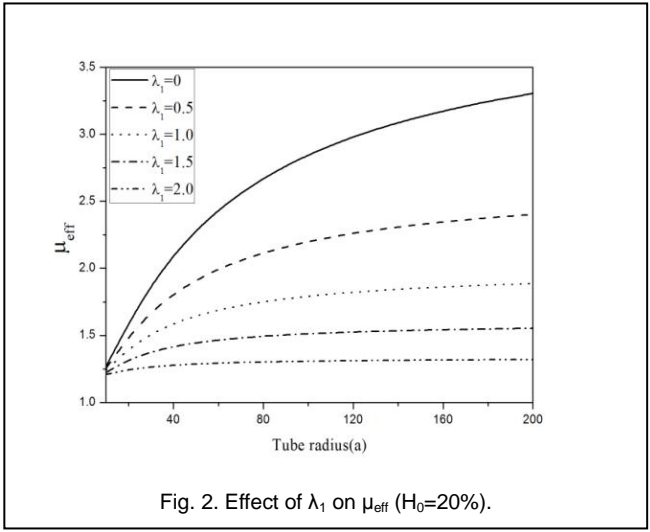

Fig. 2. Effect of $\lambda_1$ on $\mu_{eff}$ ($H_0$=20%).

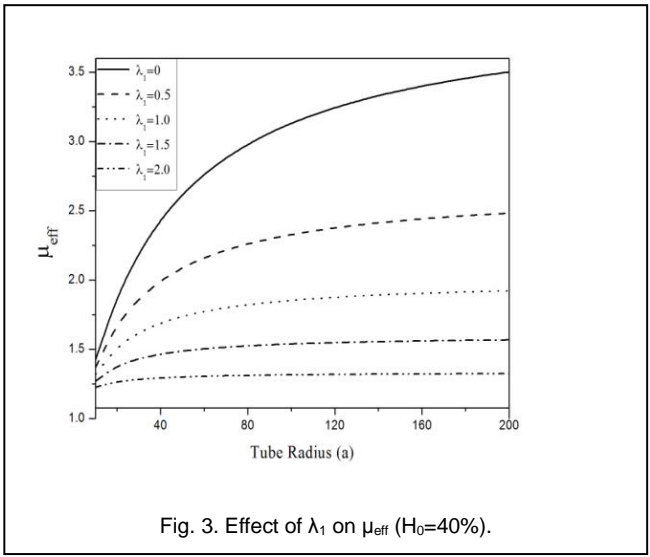

Fig. 3. Effect of $\lambda_1$ on $\mu_{eff}$ ($H_0$=40%).

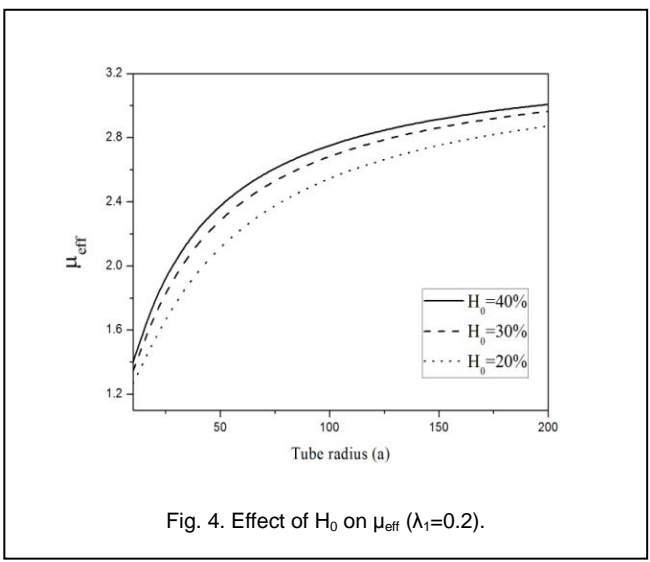

Fig. 4. Effect of $H_0$ on $\mu_{eff}$ ($\lambda_1$=0.2).

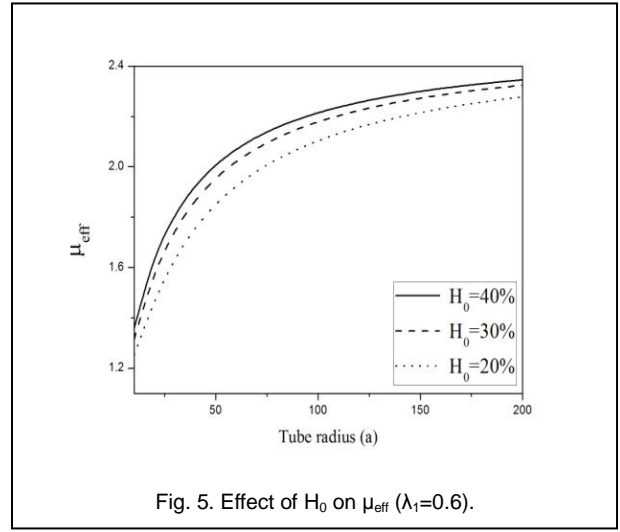

Fig. 5. Effect of $H_0$ on $\mu_{eff}$ ($\lambda_1$=0.6).

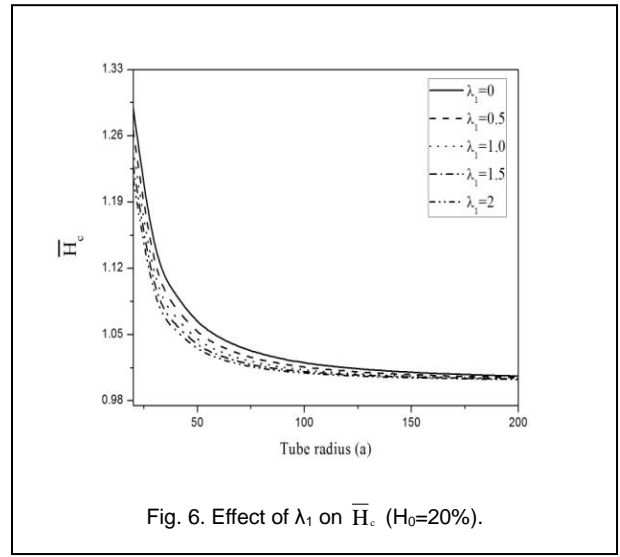

Fig. 6. Effect of $\lambda_1$ on $\overline{H}_c$ ($H_0$=20%).

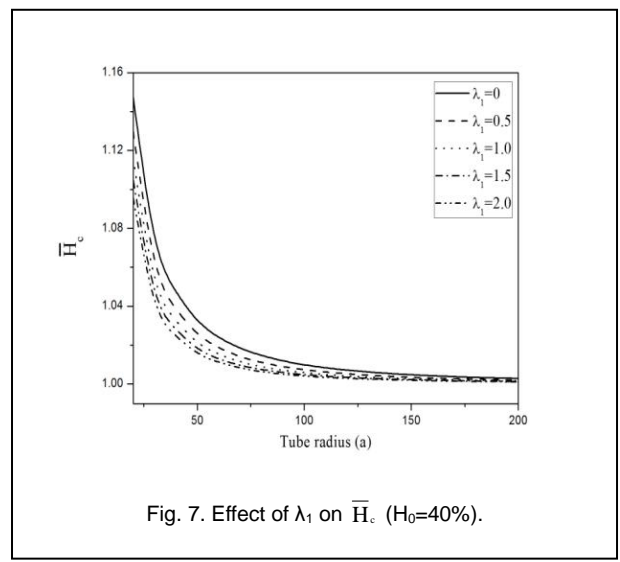

Fig. 7. Effect of $\lambda_1$ on $\overline{H}_c$ ($H_0$=40%).

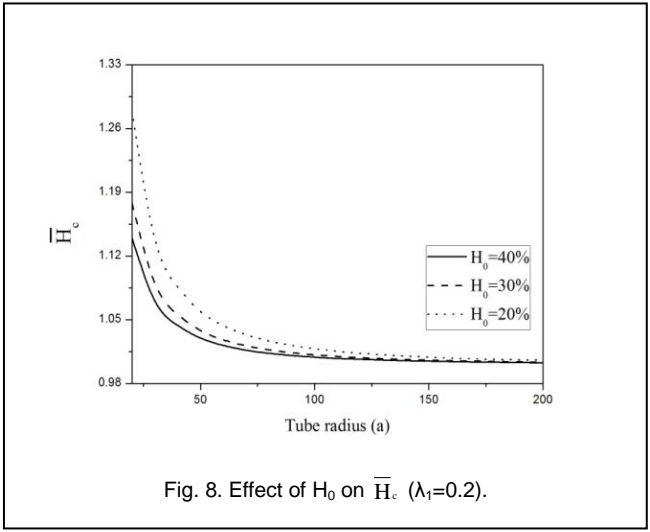

Fig. 8. Effect of $H_0$ on $\overline{H}_c$ ($\lambda_1$=0.2).

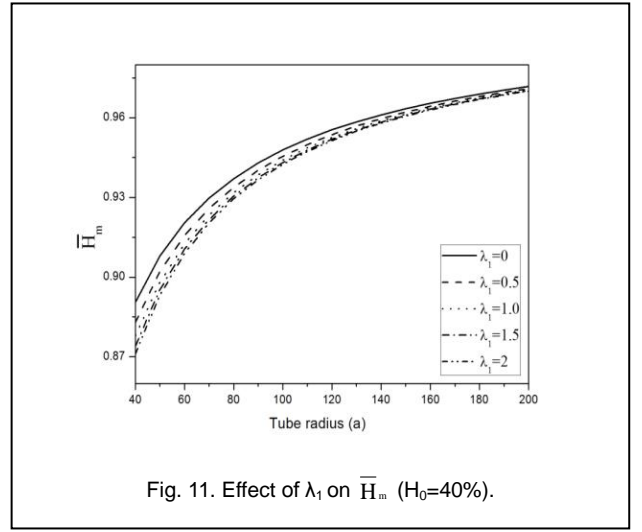

Fig. 11. Effect of $\lambda_1$ on $\overline{H}_m$ ($H_0$=40%).

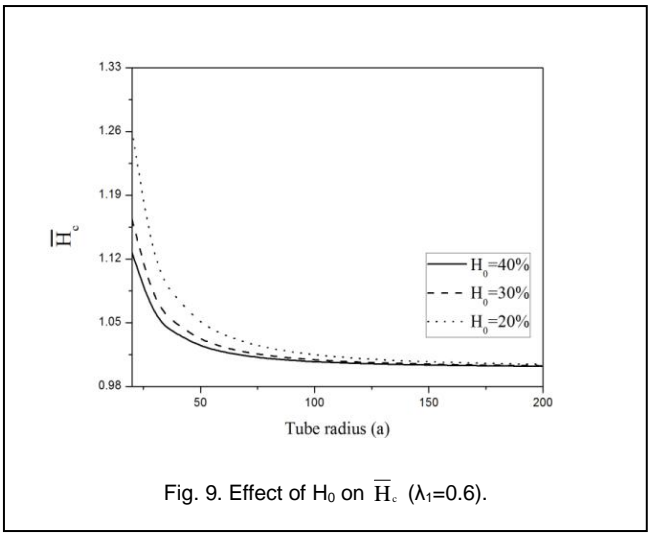

Fig. 9. Effect of $H_0$ on $\overline{H}_c$ ($\lambda_1$=0.6).

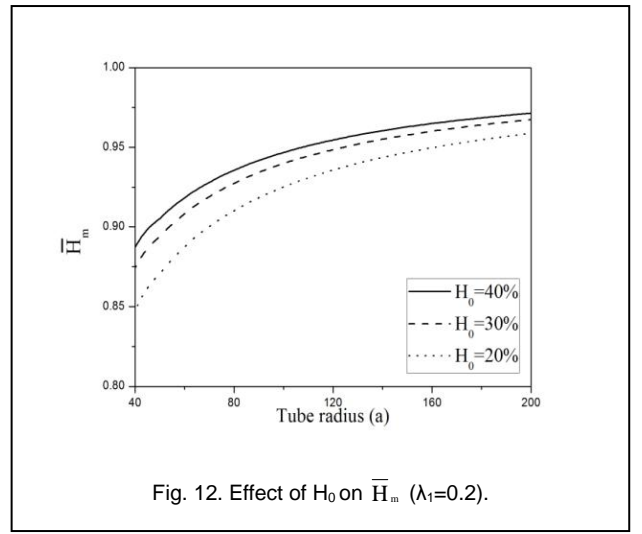

Fig. 12. Effect of $H_0$ on $\overline{H}_m$ ($\lambda_1$=0.2).

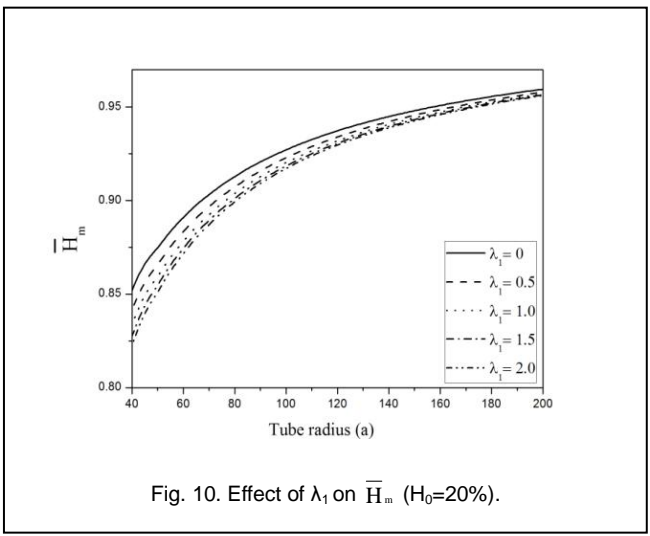

Fig. 10. Effect of $\lambda_1$ on $\overline{H}_m$ ($H_0$=20%).

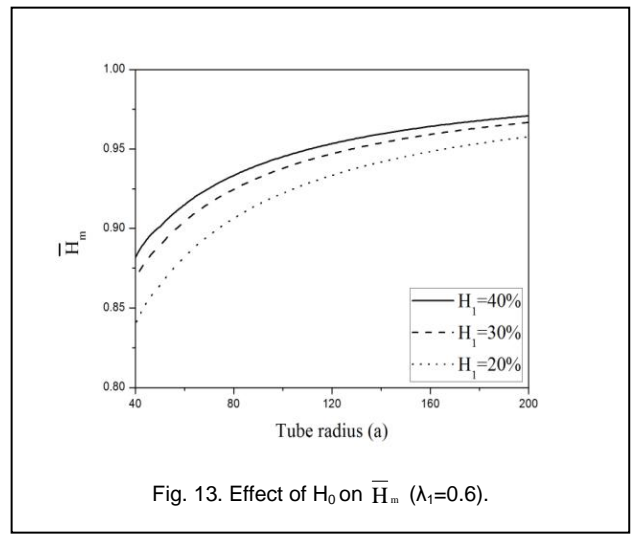

Fig. 13. Effect of $H_0$ on $\overline{H}_m$ ($\lambda_1$=0.6).

## 4 CONCLUSION

In the present paper, a two-fluid model has been proposed to describe fluid flow in small diameter tubes consisting of a Jeffrey fluid in the core region and Newtonian fluid in the peripheral region. Analytical expressions for effective viscosity, core hematocrit and mean hematocrit are obtained. It is found that the effective viscosity decreases as the Jeffrey parameter increases but increases with tube hematocrit and tube radius. Further, the core hematocrit decreases with Jeffrey parameter, tube hematocrit and tube radius. The mean hematocrit decreases as Jeffrey parameter increases but increases with tube hematocrit and tube radius. It is also noticed that the flow exhibits the anomalous Fahraeus-Lindquist effect.